\documentclass[submission, Phys]{SciPost}

\hypersetup{
    colorlinks,
    linkcolor={red!50!black},
    citecolor={blue!50!black},
    urlcolor={blue!80!black}
}

\usepackage[bitstream-charter]{mathdesign}
\DeclareSymbolFont{usualmathcal}{OMS}{cmsy}{m}{n}
\DeclareSymbolFontAlphabet{\mathcal}{usualmathcal}

\usepackage{graphicx}  
\usepackage{booktabs}
\usepackage{upgreek}
\usepackage{url}
\usepackage{physics}
\usepackage{xpatch}
\usepackage{xspace}
\usepackage{caption}
\usepackage[list=true,
labelformat=brace, position=top]{subcaption}
\usepackage[separate-uncertainty=true]{siunitx}
\usepackage{hepnames}
\usepackage{xpatch}
\usepackage{xspace}

\makeatletter
\xpatchcmd\@HepConStyle
 {\edef\@upcode{\updefault}}
 {\ifdefined\shapedefault\edef\@upcode{\shapedefault}\else\edef\@upcode{\updefault}\fi}
 {}{}
\makeatother

\newcommand{\threepi}{\ensuremath{\Ppiminus\Ppiplus\Ppiminus\Pnut}\xspace}
\newcommand{\tauthreepi}{\ensuremath{\Ptauon\to\threepi}\xspace}
\newcommand{\tautothreepi}{\tauthreepi}

\newcommand{\eeqq}{\ensuremath{\APelectron\Pelectron\to\Pquark\APquark}\xspace}
\newcommand{\eett}{\ensuremath{\APelectron\Pelectron\to\APtauon\Ptauon}\xspace}

\newcommand{\aots}{\ensuremath{\mathrm{a}_1(1260)}\xspace}

\newcommand{\aoft}{\ensuremath{\mathrm{a}_1(1420)}\xspace}
\newcommand{\rhoss}{\ensuremath{\Prho(770)}\xspace}
\newcommand{\fnotfh}{\ensuremath{\mathrm{f}_0(500)}\xspace}
\newcommand{\fnotne}{\ensuremath{\mathrm{f}_0(980)}\xspace}
\newcommand{\fnotfth}{\ensuremath{\mathrm{f}_0(1500)}\xspace}
\newcommand{\ft}{\ensuremath{\mathrm{f}_2(1270)}\xspace}
\newcommand{\rhoprime}{\ensuremath{\Prho(1450)}\xspace}
\newcommand{\rhosh}{\ensuremath{\Prho(1700)}\xspace}
\newcommand{\omegase}{\ensuremath{\Pomega(782)}\xspace}

\newcommand{\WAVE}[3]{\ensuremath{#1 \, [#2 \, \Ppi]_{\mathrm{#3}}}\xspace}

\newcommand{\pipiS}{\ensuremath{(\Ppi\Ppi)_{\mathrm{S}}}\xspace}
\newcommand{\pipiP}{\ensuremath{(\Ppi\Ppi)_{\mathrm{P}}}\xspace}
\newcommand{\ApipiS}{\WAVE{1^+}{\pipiS}{P}}
\newcommand{\VpipiP}{\WAVE{1^-}{\pipiP}{P}}

\newcommand{\sigmapiP}{\WAVE{1^+}{\sigma}{P}}
\newcommand{\rhopiS}{\WAVE{1^+}{\Prho(770)}{S}}

\newcommand{\omegapiP}{\WAVE{1^-}{\Pomega(782)}{P}}

\newcommand{\PWA}{PWA\xspace}
\newcommand{\QMIPWA}{QMIPWA\xspace}
\newcommand{\BDT}{BDT\xspace}

\newcommand{\rhopiSFF}{\ensuremath{(76.42 \pm 0.05 \pm 3.29)\%}\xspace}
\newcommand{\sigmapiPFF}{\ensuremath{(8.40 \pm 0.02 \pm 1.16)\%}\xspace}
\newcommand{\omegapiPFF}{\num{2.95 \pm 0.04e-03}\xspace}

\begin{document}

\begin{center}
  {\Large
    \textbf{
      Measurement of the presence of \aoft and \omegase in
      \tautothreepi at Belle\\
    }
  }
\end{center}

\begin{center}
A.~Rabusov\textsuperscript{1$\star$},
D.~Greenwald\textsuperscript{1}, and
S.~Paul\textsuperscript{1}
\end{center}

\begin{center}
{\bf 1} Technical University of Munich, Munich, Germany
\\
* a.rabusov@tum.de
\end{center}

\begin{center}
\today
\end{center}


\definecolor{palegray}{gray}{0.95}
\begin{center}
\colorbox{palegray}{
  \begin{tabular}{rr}
  \begin{minipage}{0.1\textwidth}
    \includegraphics[width=30mm]{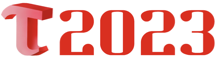}
  \end{minipage}
  &
  \begin{minipage}{0.81\textwidth}
    \begin{center}
    {\it The 17th International Workshop on Tau Lepton Physics}\\
    {\it Louisville, USA, 4-8 December 2023} \\
    \doi{10.21468/SciPostPhysProc.?}\\
    \end{center}
  \end{minipage}
\end{tabular}
}
\end{center}

\section*{Abstract}
{\bf

  We present preliminary results of a partial-wave analysis of
  \tauthreepi using data from the Belle experiment at the KEKB
  $\APelectron\Pelectron$ collider.
  We see the \aoft and a
  G-parity-violating \omegapiP wave in tauon decays.  Our results will
  improve models used in simulation studies necessary for measuring
  the electric and magnetic dipole moments and Michel parameters of
  the~\Ptau.

}

\section{Introduction}

Many studies of spin correlation in \eett, such as measuring the
electric and magnetic dipole moments of the \Ptau, analyze tauon decay
to \threepi~\cite{b2pb}.\footnote{Inclusion of charge-conjugated
decays is assumed throughout.} Lack of knowledge about the resonant
structure in \tautothreepi limits the precision of such measurements.

This decay proceeds predominantly through \aots, a broad unflavored
ground-state axial-vector meson~\cite{pluto}, whose resonance shape is
poorly known~\cite{compass-a1,cleo-ii-tau3pi,pdg2022}. What other
resonances are present and in what amounts are also poorly known. The
COMPASS and VES experiments observed the \aoft, potentially a narrow
unflavored axial-vector meson, in pion-proton
scattering~\cite{compass-a1-1420, phipsi-2019}. Seeing it in
\tauthreepi and measuring how present it is in the decay will clarify
whether it is a particle or an artifact of $\PKstar\PK$
scattering~\cite{balalaika}.

To study such matters, we perform a partial-wave analysis~(\PWA) of
\tautothreepi using \SI{980}{fb^{-1}} of data collected by the Belle
experiment~\cite{Belle} at the asymmetric $\APelectron\Pelectron$
collider KEKB~\cite{KEKB}.

\section{Event selection}

Using simulated data, we optimized our event selection to maximize
the statistical precision of our results without introducing significantly uneven
detection efficiency across the decay's phase space. Each event has
two hemispheres defined by the thrust axis calculated using all
detected charged particles and photons. We require there be three
charged particles in one, the signal hemisphere, and one in the other,
the tag hemisphere.

We use a boosted decision tree~(\BDT), implemented with ROOT's TMVA
software~\cite{tmva}, to remove events not coming from \eett. It is
trained on simulated data and bases its decision on the sum of the
momenta of charged particles and photons, the sum of energies of
charged particles, the missing mass, the cosine of the polar angle of
the missing momentum, the energy detected in the electromagnetic
calorimeter, and the event thrust; the last is the most
discriminating. All frame-dependent variables are calculated in the
$\Ppositron\Pelectron$ center-of-momentum frame.

We veto the presence of charged kaons in the signal hemisphere by
requiring the two particles with like charges be consistent with being
pions. We veto the presence of neutral kaons by requiring the mass of
each pair of oppositely charged pions in the signal hemisphere be more
than \SI{12}{MeV} from the known \PKzero mass~\cite{pdg2022}. And we
reduce the presence of neutral pions by requiring the sum of photon
energies in the signal hemisphere be below \SI{480}{MeV}, summing over
photons with at least \SI{40}{MeV}. The last requirement reduces the
ratio of the number of events with an additional \Ppizero to the number
of signal events from $51$\% to $15$\%.

We find \num{55e6} events, with 82\% purity and 32\% efficiency to
find signal---the largest sample of \tautothreepi yet
analyzed. Background events come mostly from \eeqq, with $\Pquark =
\Pup, \Pdown, \Pstrange, \Pcharm$, and from \eett with the \Ptau in
the signal hemisphere decaying to
$\Ppiminus\Ppiplus\Ppiminus\Ppizero$. We use a neural network to model
the background in our partial-wave analysis;
see~\cite{polu-nn,ichep2022} for more details.

\section{Partial-wave analysis}

The phase space of \tautothreepi has seven dimensions. We parameterize
our model intensity in the helicity angle of the \Pnut, the Euler
angles of the three-pions plane in the \Ptau rest frame, the
$\Ppiplus\Ppiminus$ squared masses, $s_1$ and $s_2$, and the mass of
the three pions, $m_{3\Ppi}$~\cite{Kuehn1995}. We average the
intensity over the Euler angle that is unmeasurable because the \Pnut
cannot be detected~\cite{ichep2022}.

We fit to the data independently in disjoint contiguous bins of
$m_{3\Ppi}$ to decompose it into partial waves using an isobar model
and the tensor formalism of~\cite{krinner-tensor}. We assume that the
decay proceeds through a resonance $\mathrm{X}^-$ that decays to three
charged pions via a sequence of two-body decays,
$\mathrm{X}^-\to\xi^0\Ppiminus$ and $\xi^0\to\Ppiminus\Ppiplus$, where
$\xi^0$ is an isobar. The only requirement on $\mathrm{X}^-$ in the
partial-wave decomposition is that its spin and parity, $J^P$, be
$0^-$, $1^+$, or $1^-$; the presence of the last would violate G
parity.

We allow $\xi^0$ to be \rhoss, \rhoprime, \fnotfh, \fnotne, \fnotfth,
\ft, or \omegase. We model them all with the relativistic Breit-Wigner
function with masses and widths the same as in the COMPASS
\PWA~\cite{compass-a1}, except for the \fnotfh, which we model with
the broad \pipiS component described in~\cite{kachaev}. Angular
momentum up to 3 is allowed between $\xi^0$ and the remaining pion. We
denote a partial wave by $J^P [\xi^0 \Ppi]_L$ for specific isobar
resonances $\xi^0$ and $J^P [(\Ppi\Ppi)_j \Ppi]_L$ for generic isobars
with spin $j$ and total angular momentum.

The preliminary results of the \PWA were presented
in~\cite{HADRON2023}. Here we present an update that includes
systematic uncertainties.  We observe that the most intense partial
wave is the \rhopiS wave, with a fit fraction of \rhopiSFF, where the
first uncertainty is statistical, and the second uncertainty is
systematic. The next most intense is the \sigmapiP wave with a fit
fraction of \sigmapiPFF. The fit fraction of a partial wave is the
integral over $m_{3\Ppi}$ of the intensity of that wave alone divided
by the same integral of the intensity of the full \PWA model. These
fractions agree with those measured by CLEO~II in
$\Ptauon\to\Ppiminus\Ppizero\Ppizero\Pnut$~\cite{cleo-ii-tau3pi}.

\begin{figure}[!t]
    \begin{center}

      \begin{subfigure}[b]{0.48\textwidth}
        \includegraphics[width=\textwidth]{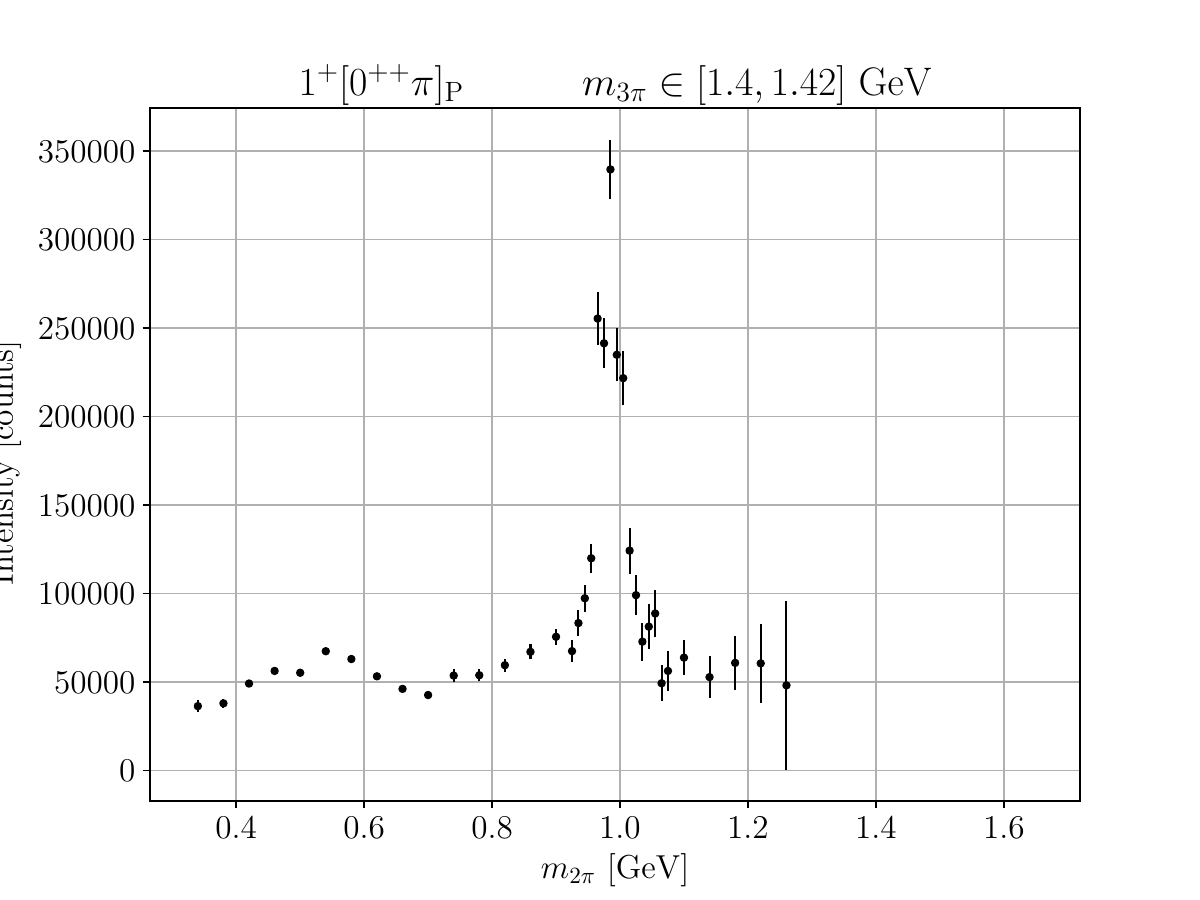}
        \caption{\ApipiS~wave.}
        \label{fig:f0_980}
      \end{subfigure}
      \begin{subfigure}[b]{0.48\textwidth}
        \includegraphics[width=\textwidth]{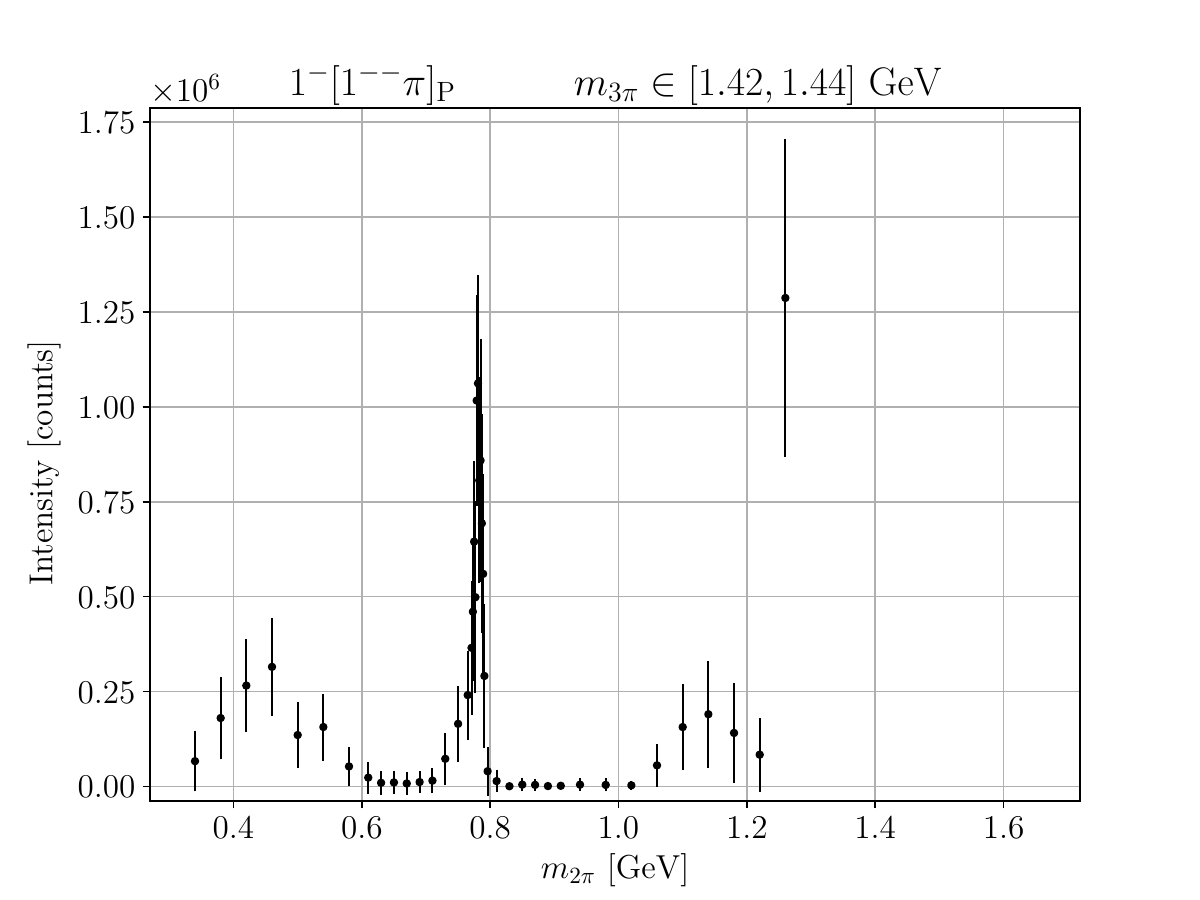}
        \caption{\VpipiP~wave.}
        \label{fig:omega}
      \end{subfigure}
      
      \caption{\QMIPWA intensities as functions of $m_{2\pi}$ with
        statistical uncertainties.}
      
    \end{center}
\end{figure}

We use quasi-model-independent \PWA~(\QMIPWA), also known as
freed-isobar \PWA,~\cite{zm} to verify our
model. We replace the \pipiS and \rhopiS models with complex step
functions, letting the fit optimize their values~\cite{HADRON2023}. We
observe the narrow peak of the \fnotne in the \pipiS wave, as shown in
Fig.~\ref{fig:f0_980}.

In~\cite{mirkes}, Mirkes and Urech stated that $1^-$ intensity in
\tautothreepi comes from the G-violating decay of
$\omegase\to\Ppiplus\Ppiminus$, where \omegase is produced by decay of
a \rhoss, \rhoprime, or \rhosh. We free the \VpipiP~wave in our
\QMIPWA and observe a narrow peak at \SI{782}{MeV}, as shown in
Fig.~\ref{fig:omega}. We include the \omegapiP~wave in the
conventional \PWA and measure a fit fraction of \omegapiPFF,
consistent with the prediction of \num{4e-3} in~\cite{mirkes}.

\section{Conclusion}

We will soon provide an updated model for \tautothreepi with about 15
partial waves and statistical and systematic uncertainties. It will be
useful for simulating \tautothreepi, necessary for measurement of the
electric and magnetic dipole moments of the \Ptau. We see the \aoft
and \omegapiP~wave in tauon decays in both conventional \PWA and
\QMIPWA. This is their first sighting in tauon decay.

\section*{Acknowledgments}

We acknowledge Florian Kaspar for providing his code to train a neural
network on simulated background data, Dmitrii Ryabchikov for providing
his code to resolve ambiguities in \QMIPWA, and Stefan Wallner for
cross checking our acceptance correction scheme.

\bibliography{TAU2023_arabusov_tau3pi_PWA.bib}

\nolinenumbers

\end{document}